# Deep Separable Spatiotemporal Learning for Fast Dynamic Cardiac MRI

Zi Wang, Min Xiao, Yirong Zhou, Chengyan Wang, Naiming Wu, Yi Li, Yiwen Gong, Shufu Chang, Yinyin Chen, Liuhong Zhu, Jianjun Zhou, Congbo Cai, He Wang, Di Guo, Guang Yang, and Xiaobo Qu

***Abstract*—Dynamic magnetic resonance imaging (MRI) plays an indispensable role in cardiac diagnosis. To enable fast imaging, the k-space data can be undersampled but the image reconstruction poses a great challenge of high-dimensional processing. This challenge necessitates extensive training data in deep learning reconstruction methods. In this work, we propose a novel and efficient approach, leveraging a dimension-reduced separable learning scheme that can perform exceptionally well even with highly limited training data. We design this new approach by incorporating spatiotemporal priors into the development of a Deep Separable Spatiotemporal Learning network (DeepSSL), which unrolls an iteration process of a 2D spatiotemporal reconstruction model with both temporal low-rankness and spatial sparsity. Intermediate outputs can also be visualized to provide insights into the network behavior and enhance interpretability. Extensive results on cardiac cine datasets demonstrate that the proposed DeepSSL surpasses state-of-the-art methods both visually and quantitatively, while reducing the demand for training cases by up to 75%. Additionally, its preliminary adaptability to unseen cardiac patients has been verified through a blind reader study conducted by experienced radiologists and cardiologists. Furthermore, DeepSSL enhances the accuracy of the downstream task of cardiac segmentation and exhibits robustness in prospectively undersampled real-time cardiac MRI.**

***Index Terms*—Deep learning, magnetic resonance imaging, dynamic imaging, image reconstruction**

## I. Introduction

DYNAMIC magnetic resonance imaging (MRI) is a non-invasive and radiation-free imaging modality that can simultaneously reveal spatial anatomical structures and temporal physiological functions [1]. It plays an indispensable role in clinical applications such as cardiac cine MRI, but suffers from the prolonged data acquisition. Accelerated imaging facilitates the achievement of high spatiotemporal resolution, improvement of patient comfort, and reduction of motion-induced artifacts [2]. For dynamic MRI acceleration, parallel imaging [3, 4] and sparse sampling [2, 5] have been widely used to provide k-t (i.e., k-space and temporal) undersampling with a high acceleration factor (AF).

Over the past two decades, to reconstruct images from undersampled data, many model-based methods have been established by exploring different signal priors to obtain promising results. Among them, sparse and low-rank priors attract the most attentions. Specifically, sparse priors are utilized for spatiotemporal data regularization in some transform domains (e.g., temporal Fourier and/or wavelet transforms [2, 6-10]). Low-rank methods commonly involve the partial separability of a Casorati matrix [5, 11] and the structured low-rankness of temporal correlations [12-14]. Moreover, based on their complementarity, some methods such as low-rank and sparse (L&S) [11, 15] and low-rank plus sparse (L+S) [16-18] have also emerged that combine two priors. Their excellent reconstruction performance prompts us for further development.

In recent years, deep learning has emerged as a powerful tool in dynamic MRI, providing high-quality images and ultra-fast reconstruction speed [19-22]. Unlike end-to-end deep networks directly learn the mapping for reconstruction [23, 24], many

This work was supported in part by the National Natural Science Foundation of China under grants 62331021, 62122064, and 62371410, Natural Science Foundation of Fujian Province of China under grants 2023J02005, 2021J011184, and 2022J011425, President Fund of Xiamen University under grant 20720220063, UKRI Future Leaders Fellowship under grant MR/V023799/1, Xiamen University Nanqiang Outstanding Talents Program, and China Scholarship Council under grant 202306310177.

Zi Wang, Yirong Zhou, Congbo Cai, and Xiaobo Qu* are with the Department of Electronic Science, Intelligent Medical Imaging R&D Center, Fujian Provincial Key Laboratory of Plasma and Magnetic Resonance, National Institute for Data Science in Health and Medicine, Xiamen University, China (*Corresponding author, email: quxiaobo@xmu.edu.cn).

Min Xiao is with the Institute of Artificial Intelligence, Xiamen University, China.

Chengyan Wang and He Wang are with the Human Phenome Institute, Fudan University, China.

Naiming Wu and Yi Li are with the Department of Imaging, Xiamen Cardiovascular Hospital of Xiamen University, School of Medicine, Xiamen University, China.

Yiwen Gong is with the Department of Cardiovascular Medicine, Heart Failure Center, Ruijin Hospital Lu Wan Branch, Shanghai Jiaotong University School of Medicine, China.

Shufu Chang is with the Shanghai Institute of Cardiovascular Diseases, Zhongshan Hospital, Fudan University, China.

Yinyin Chen is with the Department of Radiology, Zhongshan Hospital, Fudan University, Department of Medical Imaging, Shanghai Medical School, Fudan University and Shanghai Institute of Medical Imaging, China.

Liuhong Zhu and Jianjun Zhou are with the Department of Radiology, Zhongshan Hospital, Fudan University (Xiamen Branch), Fujian Province Key Clinical Specialty Construction Project (Medical Imaging Department), Xiamen Key Laboratory of Clinical Transformation of Imaging Big Data and Artificial Intelligence, China.

Di Guo is with the School of Computer and Information Engineering, Xiamen University of Technology, China.

Guang Yang is with the Department of Bioengineering and Imperial-X, Imperial College London, United Kingdom.

Zi Wang is also with the Department of Bioengineering, Imperial College London, United Kingdom.

unrolled deep networks stand out by merging the advantages of model-based and deep learning methods mentioned above [25-31]. These methods often unroll the iterative solving process of conventional models to deep networks and effectively utilize the unique MRI physical information (e.g., data consistency and coil sensitivity encoding [25-27, 32]) and/or explore the signal priors (e.g., sparsity and low-rankness [28-30]). In addition to presenting promising results, part of them also attempt to provide a certain interpretability [22].

Dynamic cardiac MRI reconstruction is a natural high-dimensional problem with high computational complexity. Up to now, most existing deep learning methods [24-30] still try to confront this complex problem directly and require numerous training data to obtain satisfactory results. However, dynamic cardiac MRI data collection is often time-consuming and highly susceptible to the cardiac/respiratory-induced motion [33-35], leading to the scarcity of available training data. Therefore, developing a new method that can effectively train an image reconstruction network with limited training data is expected.

In this work, we propose a new separable learning scheme for dynamic cardiac MRI reconstruction. Specifically, for the 2D dynamic MRI, the frequency encoding (FE) direction is always fully sampled, while the imaging acceleration happens in the phase encoding-temporal (PE-t) space by randomly skipping the PE for each temporal frame (Figs. 1(a)-(c)). Then, by taking the 1D inverse Fourier transform (IFT) along the FE, the 3D k-t data can be separate into many 2D k-t data (Fig. 1(d)-(f)). It is easy to find that all 2D k-t data have the same undersampling scenario. Therefore, instead of using the whole 3D samples in direct learning [25-28, 30], we can train a deep network on these individual 2D samples to alleviate the computational challenges.

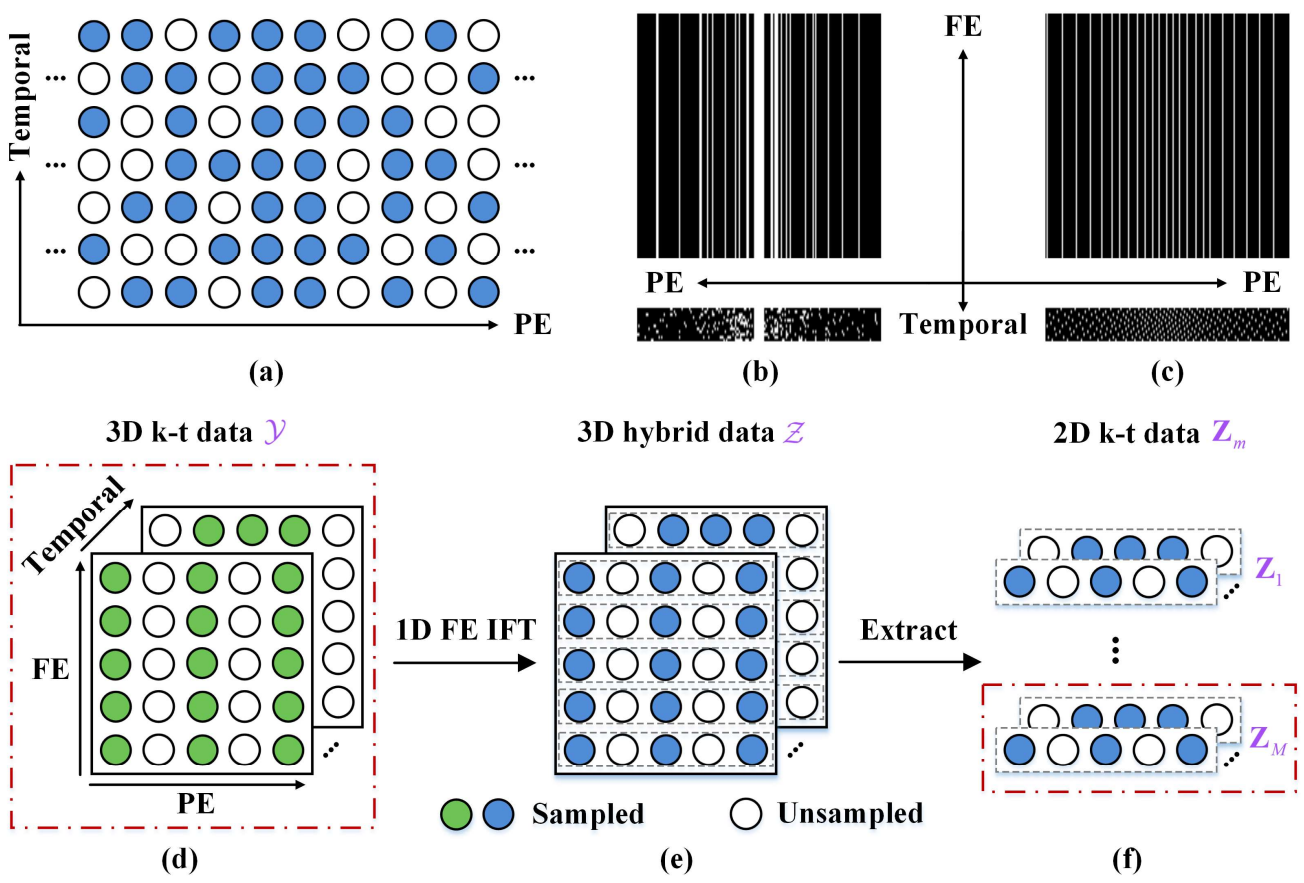

Fig. 1. Graphical illustration of the k-t undersampling and the process of reducing the scale of the reconstruction problem from 3D to 2D. (a) is the undersampling in the PE-t space. (b) is the k-t random undersampling pattern. (c) is the k-t variable density incoherent spatiotemporal acquisition (VISTA) pattern [36]. The regions marked by the red rectangle represent one of the (d) 3D and (f) 2D reconstruction, respectively Note: The proposed method can handle multi-coil data but the coil dimension is omitted here for plot simplification.

It is easy to observe that our separable learning not only reduces the dimensionality of the reconstruction problem from 3D to 2D, but also leads to the significant data enlargement with a factor equal to the dimension of FE (Table I). This nice property significantly alleviates the high dependence of deep networks on the number of training datasets. Fig. 2 illustrates an example that our separable learning is far superior to recent direct learning methods [26, 28] when training cases are highly limited (e.g., only 5 cases from *in vivo* dataset).

TABLE I

TRAINING INFORMATION FOR DIFFERENT LEARNING SCHEMES.

| Type | Problem dimensionality | Number of training samples | Variables of each training sample |
|---|---|---|---|
| Direct learning | 3D (2D spatial and 1D temporal) | $N_{TC} \times N_{slice}$ | $M \times N \times T$ |
| Separable learning | 2D (1D spatial and 1D temporal) | $N_{TC} \times N_{slice} \times M$ | $N \times T$ |

Note: $N_{TC}$ is the number of training cases. $N_{slice}$ is the number of slices each case. $M$, $N$, $T$ represent the dimension of the FE, PE, and temporal frame, respectively.

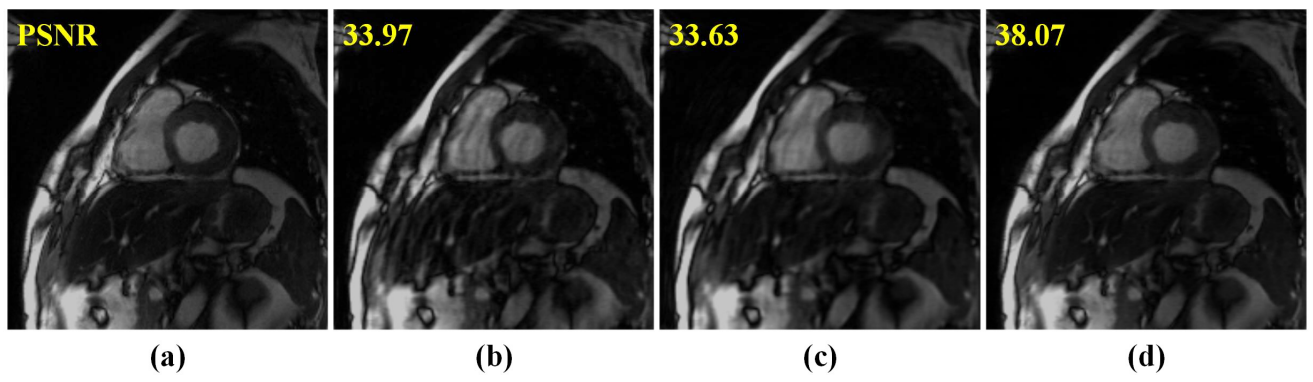

Fig. 2. A cardiac cine reconstruction example using only 5 cases from *in vivo* dataset for network training. (a) Fully sampled image. (b)-(c) are reconstructed images of two state-of-the-art direct learning methods DL-ESPIRiT [26] and SLR-Net [28], respectively. (d) is the reconstructed image of our separable learning method. Note: The k-t random undersampling pattern with AF=6 is used. PSNR(dB) is listed for each reconstruction.

Moreover, we integrate the dimension-reduced separable learning scheme with spatiotemporal priors to develop a Deep Separable Spatiotemporal Learning network (DeepSSL). It formulates both temporal low-rank and spatial sparse priors as regularized terms in an optimization model and unroll its iteration process into a deep network.

Our main contributions are summarized as follows:

1) The proposed separable learning not only reduces the problem dimensionality from 3D to 2D, but also greatly increases the number of available training samples, making network training and generalization more efficient.

2) The designed network unrolls the iteration process of a reconstruction model, which simultaneously explores the temporal low-rankness and image sparsity. We also explain our network by visualizing the intermediate results.

3) Extensive results show that, for both healthy cases and patients, the proposed DeepSSL provides superior performance under objective criteria and blind reader study. Besides, it can further benefit the downstream task of cardiac segmentation and apply in prospectively undersampled real-time cardiac MRI.

## II. Related Work

### A. Low-rank and Sparse Priors for Dynamic MRI

Low-rank and sparse modeling are powerful approaches for dynamic MRI reconstruction [2, 5]. For low-rank methods based on the partial separability of a Casorati matrix, some requires collecting navigator data to pre-estimate the temporal basis function for fix-subspace reconstruction [5], while others take it as a matrix recovery problem without navigator data [11].

Structured low-rankness of spatiotemporal correlations is also explored by constructing different structured matrices [12-14]. In addition to above global low-rank priors, local low-rank modeling can produce less temporal blurring by constraining the correlation of image patches [37]. Besides, sparse priors are also widely employed for regularization in different transform domains (e.g., temporal Fourier and/or wavelet transforms [2, 6-10]). Combining low-rank and sparse priors, many methods such as low-rank and sparse (L&S) [11, 15, 38] and low-rank plus sparse (L+S) [16-18] have emerged for improved reconstructions. More history can be found in reviews [39, 40]. However, their manual selection of regularization terms and parameters is cumbersome and not robust [41].

Introducing these priors into the design of deep learning, many unrolled deep networks stand out. Based on the ideas of L&S and L+S methods, SLR-Net [28] and L+S-Net [29] design learnable 3D sparse transforms and singular value thresholding, respectively. Sparse-inspired DUS-Net [30] introduces adaptive soft-thresholding with attention mechanism [42]. However, these deep learning methods still directly process high-dimensional data (i.e., direct learning), resulting in data-heavy training and high computational burden. Therefore, it is necessary to develop an efficient and high-performance deep learning method.

### B. Separable Reconstruction

Separable reconstruction methods have been used in MRI community to improve computational efficiency of model-based methods [2, 43, 44]. They usually employ the 1D IFT along the FE and then split the data for separable reconstructions at each FE position. Several methods perform parallel data processing and rely on autocalibration signals to estimate the coil sensitivity maps for image domain encoding [43, 44], or to estimate the weighting filters for k-space domain interpolation [45]. Many insightful sparse [2, 46, 47] and low-rank [12, 14] priors have been further used to regularize the image or k-space for improved reconstructions.

Recently, the combination of separable reconstruction and deep learning has been applied well in fast static MRI [31]. It can significantly increase the available training samples while reducing problem dimensionality, making network training and generalization easier [31]. These excellent properties arouse our interest to further develop them in dynamic MRI, which faces higher challenge to collect enough training data and has higher signal dimensionality. Compared to existing direct learning [25-28, 30], separable learning has great potential to boost the efficiency in high-dimensional applications (Table I).

## III. Proposed Method

In this section, we first formulate our basic problem on dynamic MRI and introduce the proposed dimension-reduced separable learning scheme. Then, a corresponding 2D k-t reconstruction model is presented, which simultaneously utilizes temporal low-rankness [12, 40] and spatial sparsity [2, 47]. Finally, the iteration process of a conventional reconstruction algorithm is unrolled to a deep network and its implementation details are introduced.

Our proposed deep separable spatiotemporal learning simultaneously holds the high efficiency of separable reconstructions and the high performance of prior-injected unrolled deep networks. Specifically, the separable learning scheme significantly increases the available training samples while reducing problem dimensionality, making network training and generalization easier. Meanwhile, we design learnable sparse transforms and soft-thresholding for space de-aliasing, and nonlinear null space projection for temporal artifacts elimination.

### A. Problem Formulation

For the k-t undersampling in Cartesian 2D dynamic MRI, the 3D k-t reconstruction can be split into several 2D k-t reconstructions as follows (Figs. 1(d)-(f)):

Consider the multi-coil 3D k-t data $\mathcal{Y}\in\mathbb{C}^{M\times NJ\times T}$ with non-acquired positions zero-filled and $\mathcal{U}$ is an operator that performs undersampling. $M,N,T,J$ represent the dimension of the FE, PE, temporal frame, and coil, respectively.

Taking the 1D FE IFT $\mathcal{F}_{FE}^{*}$ of $\mathcal{Y}$, we can obtain the 3D hybrid data $\mathcal{Z}\in\mathbb{C}^{M\times NJ\times T}$ (Fig. 1(e)):

$$\mathcal{Z}=\mathcal{F}_{FE}^{*}\mathcal{Y}=\mathcal{F}_{FE}^{*}\mathcal{U}\mathcal{F}_{2D}\mathcal{S}\mathcal{X}=\mathcal{U}\mathcal{F}_{PE}\mathcal{S}\mathcal{X}=\mathcal{A}\mathcal{X}\,, \tag{1}$$

where $\mathcal{X}\in\mathbb{C}^{M\times N\times T}$ is the 3D spatiotemporal image, $\mathcal{S}$ is the coil sensitivity encoding operator. $\mathcal{F}_{2D},\mathcal{F}_{PE}$ represent the 2D spatial FT and 1D PE FT, respectively. Here, for simplicity, we define the forward encoding operator as $\mathcal{A}=\mathcal{U}\mathcal{F}_{PE}\mathcal{S}$.

Given the 1D IFT decoupling, the $m^{\text{th}}$ row data $\mathbf{Z}_m\in\mathbb{C}^{NJ\times T}$ of $\mathcal{Z}$ can be treated as an independent 2D k-t signal (Fig. 1(f). Accordingly, the $m^{\text{th}}$ row data $\mathbf{X}_m\in\mathbb{C}^{N\times T}$ of $\mathcal{X}$ is treated independently, since $\mathcal{A}$ can be perform on each 2D spatiotemporal image directly. After reconstructing all $\mathbf{X}_m$, we can combine them sequentially along the FE to yield the reconstructed 3D spatiotemporal image $\hat{\mathcal{X}}\in\mathbb{C}^{M\times N\times T}$.

### B. Dimension-Reduced Separable Learning Scheme

There are two core steps for our proposed dimension-reduced separable learning scheme: Build paired 2D k-t training datasets and train the corresponding deep network.

Fig. 3 presents the overall concept of our separable learning scheme. In step one, for a fully sampled 3D k-t data (Fig. 3(a)), we undersample it and take the 1D FE IFT to obtain the undersampled 3D hybrid data (Fig. 3(c)). Then, we extract the same row of the undersampled 3D hybrid data and fully sampled 3D spatiotemporal image as a paired 2D training sample. After that, the paired 2D training dataset is built for network training, where the undersampled 2D k-t data is input (Fig. 3(d)) and the fully sampled 2D spatiotemporal image is label (Fig. 3(f)). Note that all these 2D spatiotemporal data along the FE are used for network training.

Then, in step two, we train the deep network by minimizing the loss function between reconstructed 2D spatiotemporal images (Fig. 3(e)) and labels. Eventually, the well-trained network can provide a very clear image in Fig. 3(e) and it can be applied to dynamic MRI reconstructions.

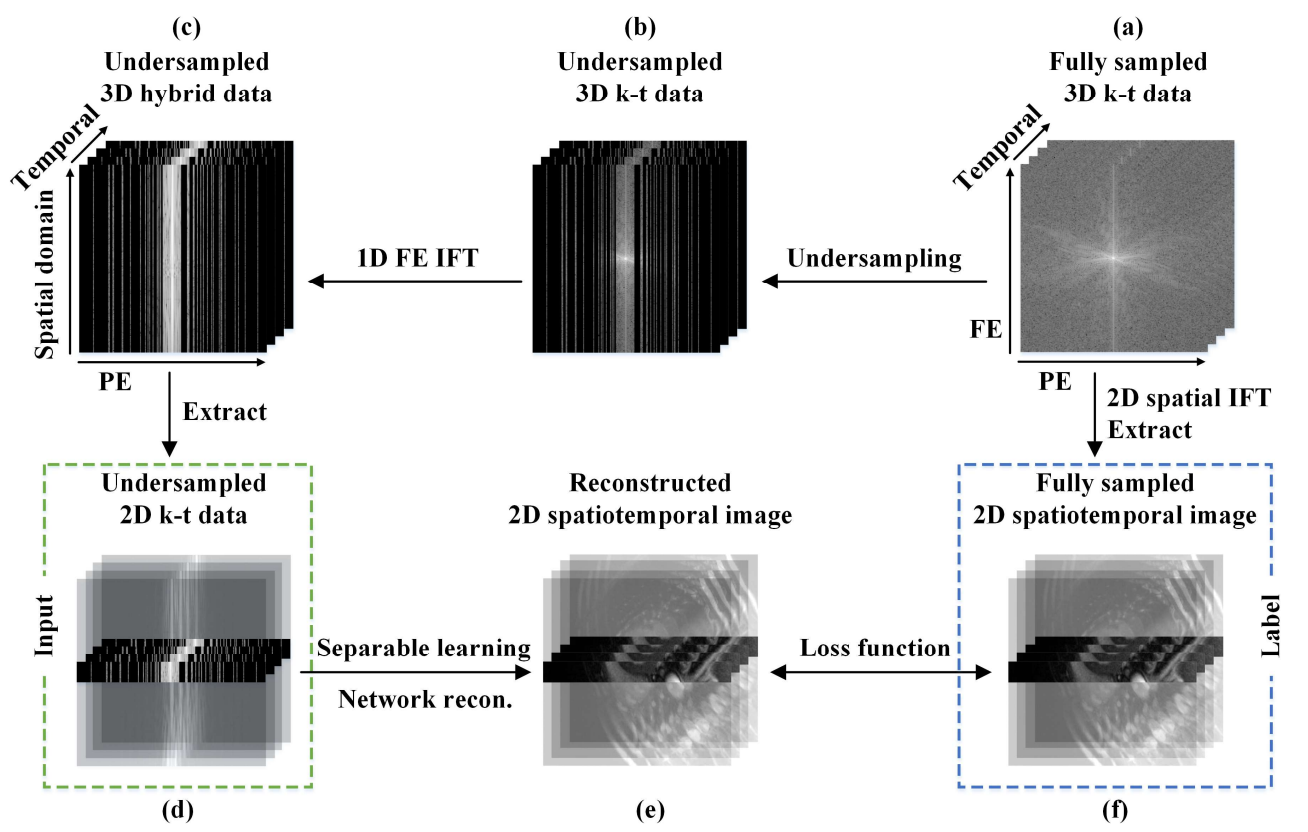

Fig. 3. Flowchart of our dimension-reduced separable learning. Note: The regions marked by the (d) green and (f) blue rectangle represent the network input and label, respectively. The coil dimension is omitted for simplification. The semitransparent images in (d)-(f) are only used for better visualization of the correspondence between them.

### C. Reconstruction Model

We build a 2D k-t reconstruction model with spatiotemporal priors. Specifically, it simultaneously utilizes the low-rankness of temporal inter-frame correlations [12, 40] and spatial sparsity in some transform domains [2, 47].

Here, the low-rank property is implicitly characterized by the null space filterbank $\mathbf{Q}$ of all temporal signals $\mathcal{E}_n\mathbf{X}_m \in \mathbb{C}^T$ [12], while the sparsity of all spatial signals $\mathcal{V}_t\mathbf{X}_m \in \mathbb{C}^N$ in some transform domains [47] is added for further improvement. The $\mathcal{E}_n$ and $\mathcal{V}_t$ denote operators that extract $n^{\text{th}}$ vector on the first dimension and $t^{\text{th}}$ vector on the second dimension from $\mathbf{X}_m$.

Thus, the 2D k-t reconstruction model with these two priors can be formulated as

$$\min_{\mathbf{X}_m} \frac{1}{2}\|\mathbf{Z}_m - \mathcal{A}\mathbf{X}_m\|_2^2 + \lambda_1 \sum_{n=1}^{N} \|\mathcal{P}(\mathbf{Q})\mathcal{E}_n\mathbf{X}_m\|_F^2 + \lambda_2 \sum_{t=1}^{T} \|\mathcal{D}\mathcal{V}_t\mathbf{X}_m\|_1, \quad (2)$$

where $\mathcal{A} = \mathcal{U}\mathcal{F}_{PE}\mathcal{S}$ is the forward encoding operator. $\mathcal{P}(\mathbf{Q})$ is the vertically cascaded Toeplitz matrix (See [31] for detailed derivations), and $\mathcal{P}(\mathbf{Q})\mathcal{E}_n\mathbf{X}_m$ represents the linear convolution between $\mathbf{Q}$ and $\mathcal{E}_n\mathbf{X}_m$. $\mathcal{D}$ is the sparse transform, such as tight frames [48, 49], $\lambda_1$ and $\lambda_2$ are regularization parameters for the temporal low-rankness term and spatial sparsity term, respectively.

To solve (2), the variable-splitting algorithm [50] is chosen to decouple the data consistency term and regularization terms. By introducing auxiliary variables $\mathbf{b}_{m,n} = \mathcal{E}_n\mathbf{X}_m$ and $\mathbf{d}_{m,t} = \mathcal{V}_t\mathbf{X}_m$, (2) can be solved by alternating three sub-problems and the $k^{\text{th}}$ iteration is

$$\mathbf{b}_{m,n}^{(k)} = \arg\min_{\mathbf{b}_{m,n}} \lambda_1 \left\|\mathcal{P}(\mathbf{Q})\mathbf{b}_{m,n})\right\|_F^2 + \frac{\mu_1}{2}\left\|\mathbf{b}_{m,n} - \mathcal{E}_n\mathbf{X}_m^{(k-1)}\right\|_2^2, \quad (3)$$

$$\mathbf{d}_{m,t}^{(k)} = \arg\min_{\mathbf{d}_{m,t}} \lambda_2 \left\|\mathcal{D}\mathbf{d}_{m,t}\right\|_1 + \frac{\mu_2}{2}\left\|\mathbf{d}_{m,t} - \mathcal{V}_t\mathbf{X}_m^{(k-1)}\right\|_2^2, \quad (4)$$

$$\mathbf{X}_m^{(k)} = \arg\min_{\mathbf{X}_m} \|\mathbf{Z}_m - \mathcal{A}\mathbf{X}_m\|_2^2 + \mu_1 \sum_{n=1}^{N} \|\mathbf{b}_{m,n}^{(k)} - \mathcal{E}_n\mathbf{X}_m\|_2^2 + \mu_2 \sum_{t=1}^{T} \|\mathbf{d}_{m,t}^{(k)} - \mathcal{V}_t\mathbf{X}_m\|_2^2. \quad (5)$$

Finally, the solutions of (3)-(5) constitutes the following iteration process of our reconstruction model:

$$\begin{cases} \mathbf{b}_{m,n}^{(k)} = \mathcal{E}_n\mathbf{X}_m^{(k-1)} - \dfrac{2\lambda_1}{\mu_1}(\mathcal{P}(\mathbf{Q}))^H \mathcal{P}(\mathbf{Q})\mathcal{E}_n\mathbf{X}_m^{(k-1)} \\ \mathbf{d}_{m,t}^{(k)} = \mathcal{D}^*[soft(\mathcal{D}\mathcal{V}_t\mathbf{X}_m^{(k-1)}; \dfrac{\lambda_2}{\mu_2})] \\ \mathbf{X}_m^{(k)} = (\mathcal{A}^*\mathcal{A} + \mu_1 \sum_{n=1}^{N} \mathcal{E}_n^*\mathcal{E}_n + \mu_2 \sum_{t=1}^{T} \mathcal{V}_t^*\mathcal{V}_t)^{-1}(\mathcal{A}^*\mathbf{Z}_m + \mu_1\mathbf{B}_m^{(k)} + \mu_2\mathbf{D}_m^{(k)}) \end{cases}, \quad (6)$$

where $\mu_1$ and $\mu_2$ are penalty parameters, the superscript $*$ is the adjoint operation, and $soft(x;\rho) = \max\{|x| - \rho\} \cdot x/|x|$ is the element-wise soft-thresholding. For simplicity, we define $\mathbf{B}_m = \sum_{n=1}^{N} \mathcal{E}_n^*\mathbf{b}_{m,n}$ and $\mathbf{D}_m = \sum_{t=1}^{T} \mathcal{V}_t^*\mathbf{d}_{m,t}$.

Conventional reconstruction algorithm iteratively updates each variable in (6) until convergence. Thus, the whole process is always time-consuming, while all regularization constraints and parameters are cumbersome and manually determined [41].

### D. Deep Separable Spatiotemporal Learning Network

In this work, we propose a Deep Separable Spatiotemporal Learning network (DeepSSL) by unrolling the iteration process in (6) to a deep network. As shown in Fig. 4, three network modules are elaborately designed, which correspond to three sub-problems in (6). They are name as deep temporal low-rank module, deep spatial sparse module, and data consistency module. Note that, all regularization and penalty parameters, sparse transform $\mathcal{D}$, and filterbank $\mathbf{Q}$ become learnable.

Specifically, our deep temporal low-rank module performs 1D convolution on each temporal signal for null space projection, then eliminate it to remove the artifacts and noise of the temporal domain. Meanwhile, in deep spatial sparse module, 1D convolution is performed on each spatial signal to achieve anti-aliasing. After repeating three modules several times, the final 2D spatiotemporal image is reconstructed. The detailed description of three modules are as follows.

#### 1) Deep Temporal Low-Rank Module

Fig. 4(c) shows the deep temporal low-rank module that corresponds to the first sub-equation of (6). $(\mathcal{P}(\mathbf{Q}))^H \mathcal{P}(\mathbf{Q})\mathcal{E}_n\mathbf{X}_m$ stands for projecting $\mathcal{E}_n\mathbf{X}_m$ to a null space to kill all meaningful signals while preserve artifacts and noise. This linear null space projection is a filterbank determined by $\mathbf{Q}$ [12, 40].

In this work, we directly use the multi-layer convolutional network to replace $(\mathcal{P}(\mathbf{Q}))^H \mathcal{P}(\mathbf{Q})$ to achieve nonlinear null space projection, as network has strong representation ability [31, 51]. Then, we subtract these artifacts and noise from input $\mathcal{E}_n\mathbf{X}_m$ for temporal de-aliasing, which is similar to a form of residual learning [52]. Thus, the deep temporal low-rank module is designed as

$$\mathbf{b}_{m,n}^{(k)} = \mathcal{E}_n\mathbf{X}_m^{(k-1)} - \mathcal{N}_1(\mathcal{E}_n\mathbf{X}_m^{(k-1)}), \quad (7)$$

where $\mathcal{N}_1$ is a multi-layer convolutional network. When $k = 1$, the initialized input $\mathbf{X}_m^{(0)} = \mathcal{A}^*\mathbf{Z}_m$ is the zero-filled data that has strong artifacts.

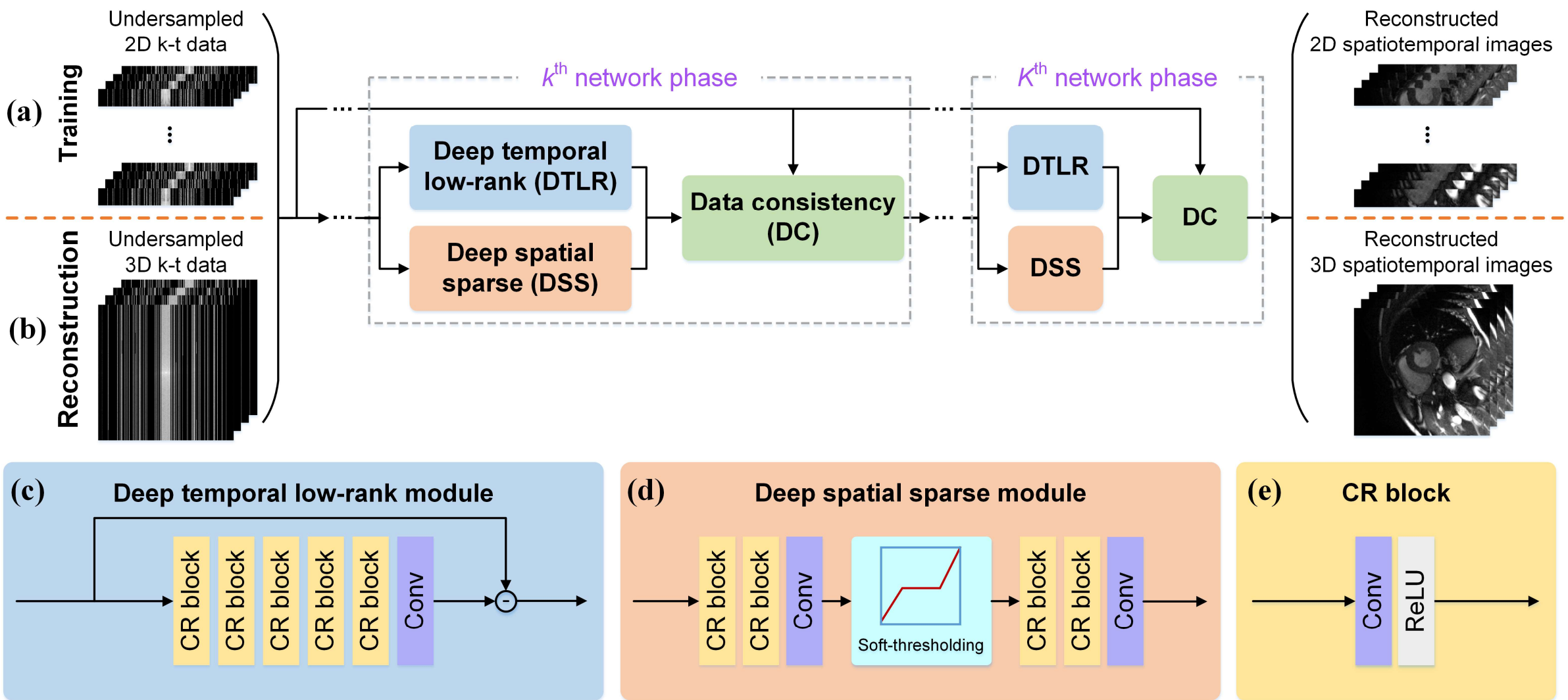

Fig. 4. The proposed DeepSSL for dynamic MRI reconstruction. This is the recursive network architecture in (a) training and (b) reconstruction stages. With the increase of the network phase, artifacts are gradually removed, and a high-quality reconstructed image can be obtained finally. (c)-(d) are the detailed structures of network modules. Note: "Conv" is the convolution. "CR" is the convolution followed by a ReLU activation.

Fig. 5 visualizes the nonlinear null space projection achieved by our designed network $\mathcal{N}_1$ and the output of this module. These observations imply that, it can successfully realize proper null space estimation to primarily preserve artifact-related signals. Moreover, with the increase of network phases, the output of $\mathcal{N}_1$ gradually approaches zero, indicating that artifacts are gradually removed and a high-quality image is yielded finally. This visualization provides a good interpretation to understand the network behavior.

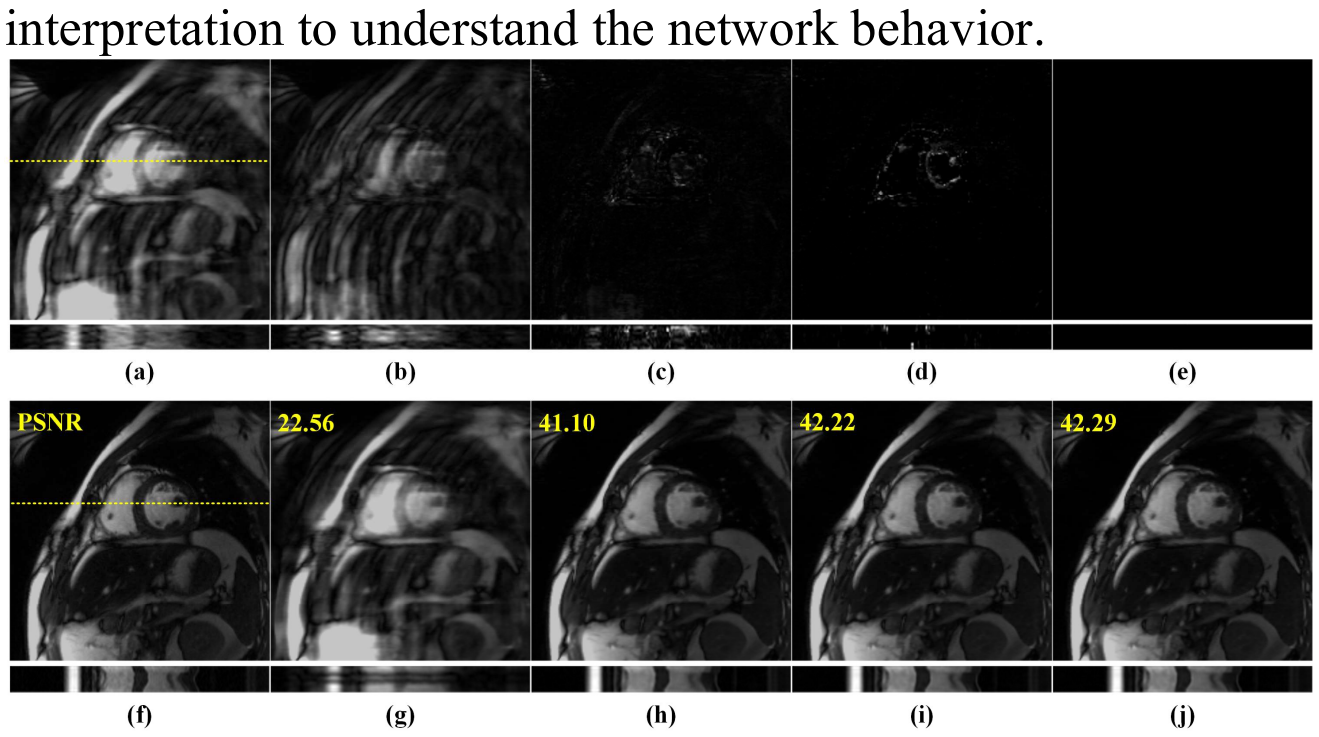

Fig. 5. The null space projections and reconstructed images of the deep temporal low-rank module at representative network phase. (a) is the zero-filled image. (b)-(e) are the null space projections at 1st, 5th, 9th, 10th (final) network phase, respectively. (f) is the fully sampled image. (g)-(j) are reconstructed images at 1st, 5th, 9th, 10th (final) network phase, respectively. Note: The k-t random undersampling pattern with AF=6 is used. The 2D spatiotemporal images are given for each reconstruction and their corresponding extracted positions are marked by the yellow dotted lines. PSNR(dB) is listed for each reconstruction.

### 2) Deep Spatial Sparse Module

Fig. 4(d) shows the deep spatial sparse module that corresponds to the second sub-equation of (6). Here, replacing the manual-craft transform $\mathcal{D}$ and threshold $\lambda_2/\mu_2$, we utilize the widely used deep thresholding network [42, 53], to learn a more general sparse transform from training datasets. Thus, the deep spatial sparse module is designed as

$$\mathbf{d}_{m,t}^{(k)} = \mathcal{N}_3[soft(\mathcal{N}_2(\mathcal{V}_t\mathbf{X}_m^{(k-1)});\theta^{(k)})], \tag{8}$$

where $\theta^{(k)}$ is the learnable threshold initialized to 0.001 and it is allowed to vary at each network phase. Note that, two multi-layer convolutional networks $\mathcal{N}_2$ and $\mathcal{N}_3$ on both sides of the soft-thresholding operator are relaxed without any invertible constraint.

### 3) Data Consistency Module

In this module, each final output is forced to maintain the data consistency to the measured data, which corresponds to the third sub-equation of (6). Here, the data consistency module is denoted more intuitively as

$$(\mathcal{F}_{PE}\mathcal{S}\mathbf{X}_m^{(k)})_p = \begin{cases} (\dfrac{\mu_1^{(k)}\mathcal{F}_{PE}\mathcal{S}\mathbf{B}_m^{(k)} + \mu_2^{(k)}\mathcal{F}_{PE}\mathcal{S}\mathbf{D}_m^{(k)}}{\mu_1^{(k)}+\mu_2^{(k)}})_p, & p \notin \Omega \\ (\dfrac{\mathbf{Z}_m + \mu_1^{(k)}\mathcal{F}_{PE}\mathcal{S}\mathbf{B}_m^{(k)} + \mu_2^{(k)}\mathcal{F}_{PE}\mathcal{S}\mathbf{D}_m^{(k)}}{1+\mu_1^{(k)}+\mu_2^{(k)}})_p, & p \in \Omega \end{cases}, \tag{9}$$

where $p$ is the index and $\Omega$ is the set of the measured positions in k-t data. $\mu_1^{(k)}$ and $\mu_2^{(k)}$ set as learnable parameters initialized to 1 and are changed at each network phase. This module implies that, at the measured positions, the data points should maintain a trade-off with $\mathbf{Z}_m$, while the update of the unmeasured data points depends entirely on the reconstruction results of other two network modules.

## E. Implementation Details

The proposed DeepSSL is an unrolled recursive network and we empirically choose $K=10$ as the overall number of network phase, due to a trade-off between reconstruction performance and time consumption. At each network phase, $\mathcal{N}_1$, $\mathcal{N}_2$, and $\mathcal{N}_3$ consist of 6, 3, and 3 convolutional layers, respectively. Each convolution layer contains 48 1D convolution filters of size 3, followed by a ReLU as activation function. Note that, the last convolution layer of each module has only 2 filters, corresponding to two channels of real and imaginary parts of data.

In the training stage, our DeepSSL is trained for 50 epochs with the Adam optimizer. Its initial learning rate is set to 0.001 with an exponential decay of 0.99, while the batch size is 64. The loss function is defined as:

$$\mathcal{L}(\mathbf{\Theta}) = \frac{1}{KC}\sum_{k=1}^{K}\sum_{c=1}^{C}\left\|\mathbf{X}_m^{ref,c} - \mathbf{X}_m^{(k),c}\right\|_2^2, \quad (10)$$

where $C$ is the number of training samples, and $\mathbf{X}_m^{ref,c}$ is the label of the $c^{\text{th}}$ training sample. DeepSSL was implemented on a server equipped with dual Intel Xeon Silver 4210 CPUs, 256 GB RAM, and the Nvidia Tesla T4 GPU (16 GB memory) in PyTorch 1.10. The typical training took about 50 hours.

In the reconstruction stage, for given undersampled 3D k-t data, we can reconstruct them through the trained DeepSSL. As shown in Fig. 4(b), the 1D FE IFT is first performed on the undersampled k-space to obtain $\mathcal{Z} = \mathcal{F}_{FE}^{*}\mathcal{Y}$, all rows of $\mathcal{Z}$ form a batch that is then reconstructed in parallel and stitched back together to yield the final 3D spatiotemporal image $\hat{\mathcal{X}}$.

## IV. Experimental Results

### A. Datasets

Two 2D cardiac cine datasets were mainly used in this paper: A dataset of healthy cases from a public CMRxRecon dataset [34] and an in-house patient dataset from our own collection. Their study protocol was Institutional Review Board approved (MS-R23) and informed consent was obtained from volunteers before examination.

#### 1) Dataset of Healthy Cases

The dataset of healthy cases was acquired using a segmented TrueFISP sequence with breath-hold at a 3T MRI scanner (Siemens MAGNETOM Vida) [34]. The cardiac cycle was segmented into 12~25 phases with a temporal resolution 50 ms according to the heart rate. Other scan parameters: Spatial resolution 2.0×2.0 mm$^2$ and slice thickness 8.0 mm. The collected images include a short-axis (SAX) view, and three long-axis (LAX) views including two/three/four-chamber (2CH/3CH/4CH) views.

There are 120 cases and each case has 10 SAX slices of size 192×192 with 12 temporal frames, and 3 LAX slices (a single slice was acquired for each 2CH/3CH/4CH view) of 192×192 with 12 temporal frames. Here, 100 cases were used for training, 10 for validation, and the remaining 10 for test.

#### 2) Datasets of Patients

A patient dataset with hypertrophic cardiomyopathy was acquired using a segmented TrueFISP sequence with breath-hold at a 3T MRI scanner (Siemens MAGNETOM Vida). The cardiac cycle was segmented into 12~25 phases with a temporal resolution 50 ms according to the heart rate. Other scan parameters: TR/TE=3.6/1.6 ms, spatial resolution 1.5×1.5 mm$^2$, and slice thickness 8.0 mm. The collected images include a SAX view, and three LAX views including 2CH/3CH/4CH views.

There are 5 cases and each case has 10 SAX slices of size 246×246 with 12 temporal frames, and 3 LAX slices (a single slice was acquired for each 2CH/3CH/4CH view) of size 224×204 with 12 temporal frames. All 5 cases were used for test.

All datasets were fully sampled k-space, and they were first retrospectively undersampled, then used for training and test. The number of coils is 10 in all two datasets. Coil sensitivity maps were pre-estimated using ESPIRiT [54].

### B. Compared Methods and Evaluation Criteria

For comparative study, a conventional method with low-rank and sparse priors L+S [17] was used as the reconstruction baseline. We also compared the proposed DeepSSL with five state-of-the-art unrolled deep learning methods including DCCNN [25], DL-ESPIRiT [26], CINE-Net [27], SLR-Net [28], and DUS-Net [30]. The training schemes of them are direct learning with 3D k-t data, which is different to our proposed separable learning with 2D k-t data. DCCNN introduce the data consistency in the 3D convolutional network, while DL-ESPIRiT and CINE-Net further utilizes the 2D spatial convolution and 1D temporal convolution. SLR-Net is a deep sparse and low-rank network with the 3D convolution and singular value decomposition. DUS-Net uses 3D convolution and introduces adaptive soft-thresholding with attention mechanism. All deep learning methods were executed according to the typical setting mentioned by the authors.

Notably, by virtue of our separable learning and 1D convolution, the proposed DeepSSL is memory-efficient. It has few network parameters (564520), about 65% of DCCNN, 61% of DL-ESPIRiT, 62% of CINE-Net, 42% of SLR-Net, and 33% of DUS-Net.

All methods were implemented on a server equipped with dual Intel Xeon Silver 4210 CPUs, 256 GB RAM, and the Nvidia Tesla T4 GPU (16 GB memory). The typical 3D k-t data reconstruction times per slice: L+S (38.78 seconds), DCCNN (1.18 s), DL-ESPIRiT (1.09 s), CINE-Net (1.08 s), SLR-Net (2.08 s), DUS-Net (2.38 s), and DeepSSL (1.43 s).

To quantitatively evaluate the image reconstruction performance, we utilized three objective evaluation criteria, including the relative $l_2$ norm error (RLNE) [55], peak signal-to-noise ratio (PSNR), and structural similarity index (SSIM) [56]. A lower RLNE, higher PSNR, and higher SSIM indicate the lower reconstruction error, less image distortions, and better details preservations, respectively. Significance of differences between different methods were further evaluated with the Wilcoxon signed-rank test.

### C. Separable versus Direct Learning

Here, we trained networks on a cardiac cine dataset of varied number of training cases $N_{TC}$, to compare the reconstruction performance of our separable learning (DeepSSL) with its direct learning version (DeepDSL). Their network architectures are only differed in the size of spatial convolution filters (1D in DeepSSL versus 2D in DeepDSL), to eliminate interferences of the network design.

Fig. 6 shows that, the separable learning significantly outperforms direct learning under small number of training cases ($N_{TC} \leq 10$). If number of training cases is extremely small ($N_{TC}$ = 5), the result of direct learning has obvious artifacts and details loss (Fig. 6(e)). With the increase of training cases, both separable and direct learning gradually improves the reconstruction. The latter improves the performance faster than the former and starts to show its advantages over the former in SSIM when $N_{TC} \geq 25$, although the former consistently performs better in RLNE and PSNR. Finally, when the number of training cases is relatively large ($N_{TC}$ = 100), direct learning achieves comparative results with separable ones in RLNE and PSNR and outperforms it in SSIM. This phenomenon may be due to the lack of spatial smoothness constraints along the FE in separable learning, leading to a loss in image similarity [31, 46]. Notably, direct learning has much more network parameters (two times that of separable learning) and costs much longer training time (90 hours versus 50 hours of separable learning).

These results imply that the proposed separable learning is a more efficient choice, since it is suitable for limited available training cases and simplifies computation.

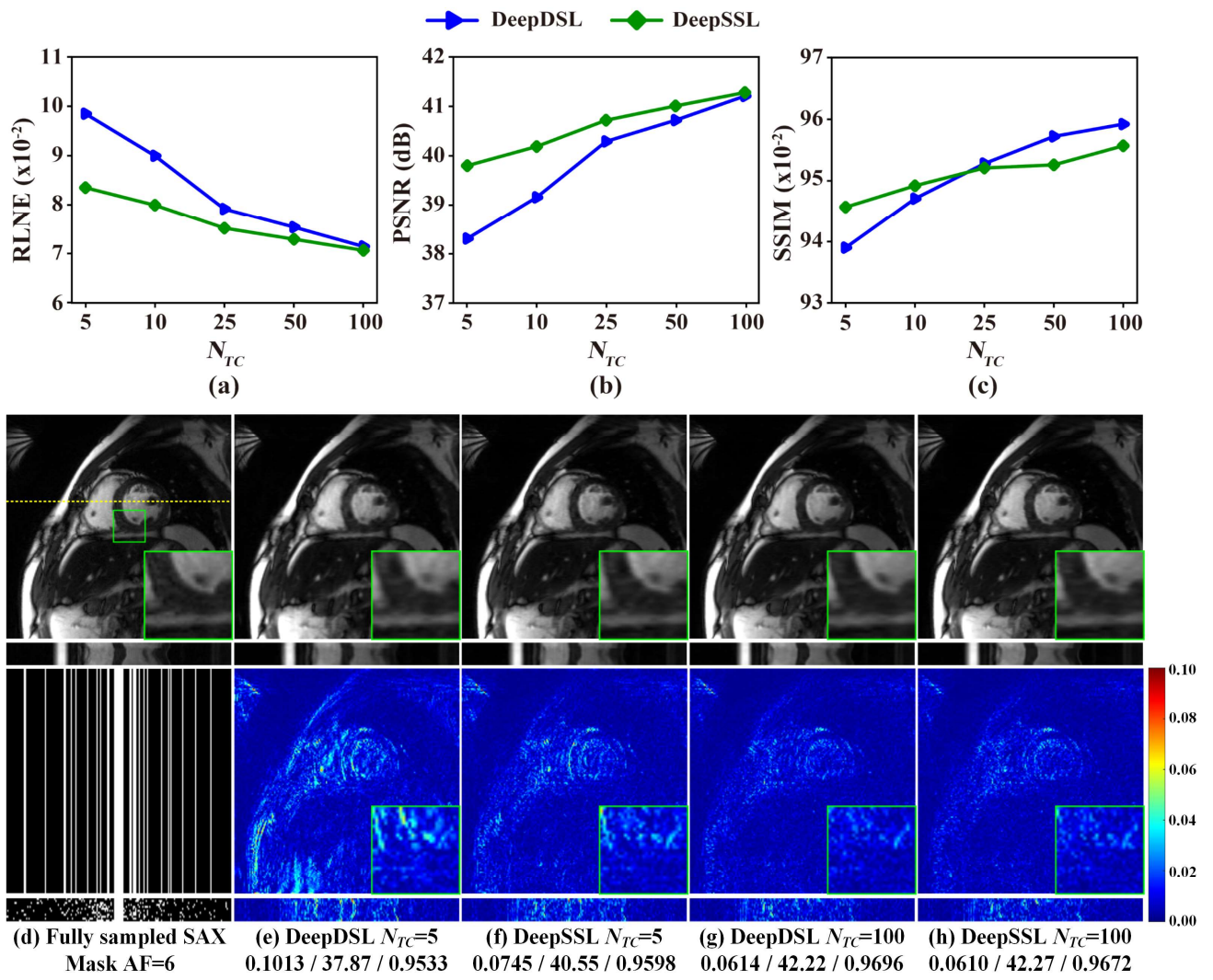

Fig. 6. Quantitative comparison of reconstructions of direct (DeepDSL) and separable (DeepSSL) learning using different number of training cases $N_{TC}$ of a cardiac cine dataset. (a)-(c) are the mean of RLNE, PSNR, SSIM of the reconstructions of the SAX view, respectively. (d) is the fully sampled image with the green zoom-in window. (e)-(f) (or (g)-(h)) are reconstructed images and the corresponding error maps using 5 (or 100) training cases with zoom-in details. Note: The k-t random undersampling pattern with AF=6 is used. The means are computed over all 10 test data of healthy cases. The 2D spatiotemporal images are given for each reconstruction and their corresponding extracted position is marked by the yellow dotted line. RLNE/PSNR(dB)/SSIM are listed for each reconstruction.

### D. Comparison with State-of-the-art Methods

Here, we trained networks on a cardiac cine dataset of varied number of training cases $N_{TC}$, to compare the reconstruction performance of our separable learning with other state-of-the-art direct learning methods.

Figs. 7(a)-(c) show that our separable learning DeepSSL consistently outperforms other direct learning networks in terms of RLNE, PSNR, and SSIM. Even with a highly limited number of training cases ($N_{TC} \leq 10$), DeepSSL already provides far superior reconstructions than the baseline L+S. In contrast, direct learning DCCNN, DL-ESPIRiT, CINE-Net, SLR-Net, and DUS-Net start to surpass the baseline L+S when $N_{TC} \geq 25$. With the increase of training cases, all methods gradually yield improvement. Finally, when a relatively large number of cases is used for training ($N_{TC}$ = 100), SLR-Net and DUS-Net achieve results comparable to our DeepSSL in terms of SSIM. The similar phenomena can also be found in the reconstruction of the LAX view (Figs. 7(d)-(f)). Notably, we observed that the proposed DeepSSL can maintain the similar performance to state-of-the-art methods, while reducing the demand for training cases by up to 75% (from 100 to 25 cases).

Representative reconstructions using 100 training cases are shown in Fig. 8. L+S, DCCNN, and DL-ESPIRiT yield results exhibiting obvious spatial and temporal artifacts. CINE-Net, SLR-Net, and DUS-Net are much better but still appear inaccurate edge brightness and cardiac details loss. The proposed DeepSSL outperforms other state-of-the-art methods in terms of three evaluation criteria, indicating the good ability of artifacts suppression and details preservation.

These results imply that, even with highly limited training data, the proposed method can work efficiently and provide state-of-the-art reconstructions of different cardiac views.

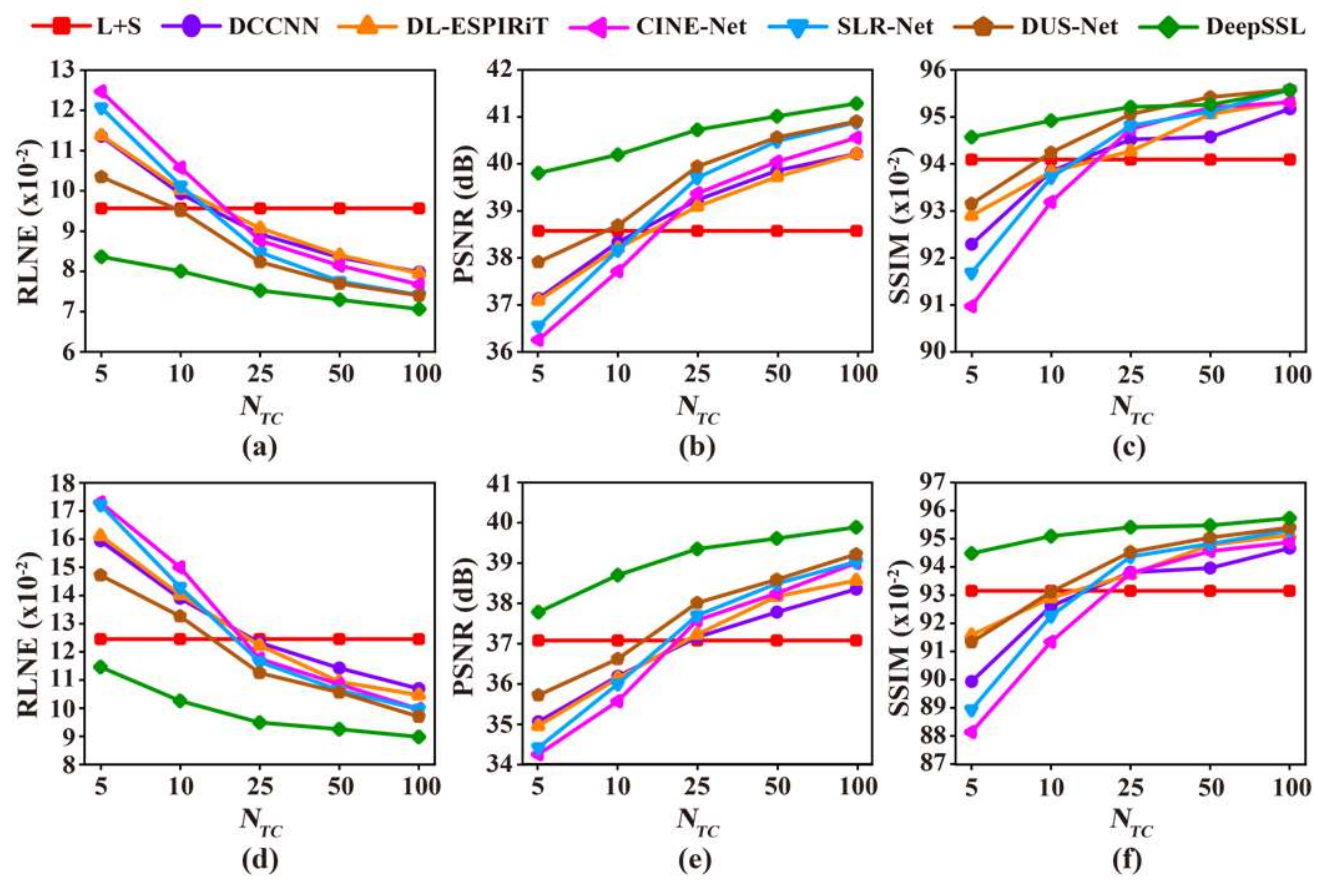

Fig. 7. Quantitative comparison of reconstructions using different number of training cases $N_{TC}$ of a cardiac cine dataset. (a)-(c) are the mean of RLNE, PSNR, SSIM of the reconstructions of the SAX view, respectively. (d)-(f) are the mean of RLNE, PSNR, SSIM of the reconstructions of the LAX view, respectively. Note: The k-t random undersampling pattern with AF=6 is used. The means are computed over all 10 test data of healthy cases.

### E. Adaptability to Unseen Patients and Reader Study

Considering that patient data often differs from healthy data, we further examined the adaptability of our method to patients, which is essential in clinical diagnosis. Here, we used a CMRxRecon dataset [34] of 100 healthy cases to train all deep learning methods, and then reconstructed an unseen cardiac cine dataset of 5 patients with hypertrophic cardiomyopathy.

Since objective criteria (e.g., RLNE, PSNR, and SSIM) might not comprehensively reflect image quality, we also conducted a reader study online through our CloudBrain platform [57] (https://csrc.xmu.edu.cn/CloudBrain.html). Six readers were invited (4 radiologists with 3/8/10/12 years' experience and 2 cardiologists with 10/12 years' experience), to independently evaluate reconstructed images from a diagnostic perspective. They were blind to all reconstruction methods, while fully sampled images were provided for reference. Three clinical-concerned criteria including signal-to-noise ratio (SNR), artifacts suppression, and overall image quality were used. Using a 5-point scale, each criterion ranged from 0 to 5 with a precision of 0.1 (i.e., 0~1: Non-diagnostic; 1~2: Poor; 2~3: Adequate; 3~4: Good; 4~5: Excellent).

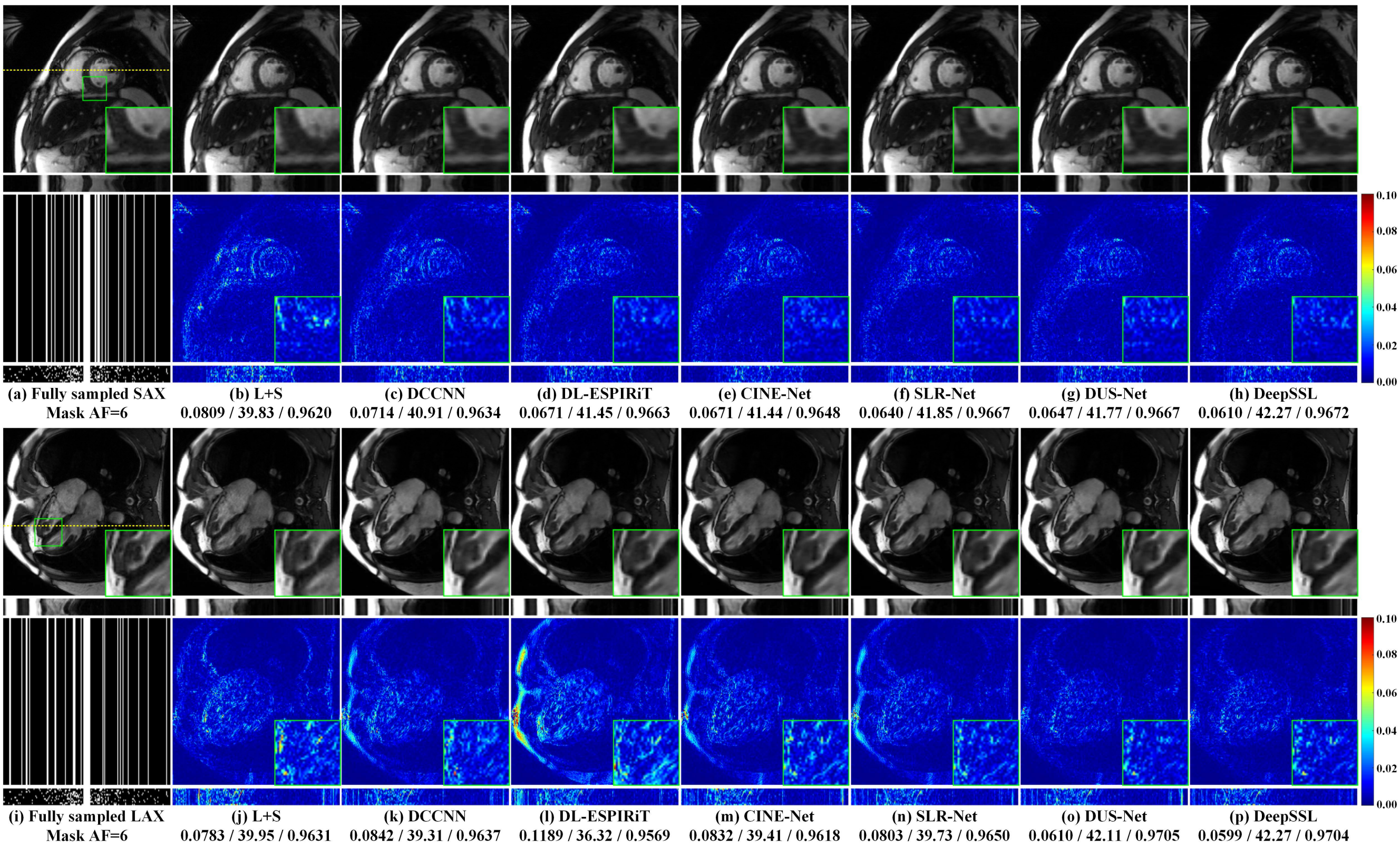

Fig. 8. Reconstruction results of a cardiac cine dataset of SAX and LAX views using different methods. (a) (or (i)) is the fully sampled image with the green zoom-in window and the k-t random undersampling pattern with AF=6. (b)-(h) (or (j)-(p)) are reconstructed images and the corresponding error maps with zoom-in details. Note: The training cases is 100. The 2D spatiotemporal images are given for each reconstruction and their corresponding extracted position is marked by the yellow dotted line. RLNE/PSNR(dB)/SSIM are listed for each reconstruction.

TABLE II

THE SCORES OF THE READER STUDY [MEAN±STD].

| Data | Method | SNR | Artifacts suppression | Overall image quality |
|---|---|---|---|---|
| SAX view | L+S | 3.10±0.31† | 3.04±0.48† | 3.29±0.45† |
| | DL-ESPIRiT | 3.60±0.31† | 3.43±0.36† | 3.62±0.34† |
| | SLR-Net | 4.05±0.28† | 3.74±0.37† | 3.99±0.36† |
| | DeepSSL | **4.23±0.24** | **3.98±0.26** | **4.21±0.25** |
| LAX view | L+S | 2.87±0.30† | 2.99±0.51† | 3.01±0.36† |
| | DL-ESPIRiT | 3.61±0.46† | 3.39±0.60† | 3.54±0.48† |
| | SLR-Net | 3.93±0.38† | 3.65±0.56† | 3.88±0.35† |
| | DeepSSL | **4.17±0.35** | **3.94±0.56** | **4.16±0.33** |

Note: There are 120 SAX images and 120 LAX images from 5 patients used for the reader study. The means and standard deviations are computed over all images, respectively. The highest scores are bold faced. "†" means the compared method has statistically significant differences (p<0.001) compared to our DeepSSL under Wilcoxon signed-rank test.

Table II shows that the proposed DeepSSL obtains highest mean scores and is the only one which all three criteria of reader study are higher or very close to 4, indicating its image quality basically steps into an excellent level from a diagnostic perspective. Besides, the differences between DeepSSL and other compared methods are statistically significant (p<0.001). As for other compared methods, they lose details and have artifacts in the reconstructed images (Fig. 9), resulting in lower scores that can only reach the adequate or good level. These results are consistent with the superiority of DeepSSL (p<0.001) on the objective criteria (RLNE, PSNR, and SSIM) in Table III.

These results demonstrate that, from both objective and clinical-concerned subjective criteria, our technique is adaptive to unseen cardiac patients and provides reliable image reconstructions.

TABLE III

RLNE ($\times 10^{-2}$)/PSNR (dB)/SSIM ($\times 10^{-2}$) OF PATIENT RECONSTRUCTIONS UNDER K-T RANDOM UNDERSAMPLING PATTERN WITH AF=6 [MEAN±STD].

| Data | Method | RLNE | PSNR | SSIM |
|---|---|---|---|---|
| SAX view | L+S | 10.66±3.80† | 38.89±2.76† | 94.92±2.43† |
| | DL-ESPIRiT | 8.53±1.99† | 40.61±2.18† | 95.32±1.69† |
| | SLR-Net | 8.47±2.04† | 40.68±2.21† | 95.07±1.72† |
| | DeepSSL | **7.74±1.43** | **41.38±1.99** | **95.55±1.45** |
| LAX view | L+S | 13.59±4.32† | 36.62±2.29† | 93.92±1.98† |
| | DL-ESPIRiT | 10.59±2.41† | 38.61±1.67† | 94.83±1.56† |
| | SLR-Net | 10.09±2.49† | 39.07±1.97† | 94.91±1.54† |
| | DeepSSL | **9.38±1.56** | **39.60±1.86** | **95.05±1.67** |

Note: The means and standard deviations are computed over all 5 test data of patients, respectively. The lowest RLNE, highest PSNR and SSIM values are bold faced. The highest scores are bold faced. "†" means the compared method has statistically significant differences (p<0.001) compared to our DeepSSL under Wilcoxon signed-rank test.

### F. Example Downstream Task: Cardiac Segmentation

Generally, high-quality reconstructions benefit downstream tasks in image analysis, achieving close performance compared to fully sampled images. Here, we focused on the downstream

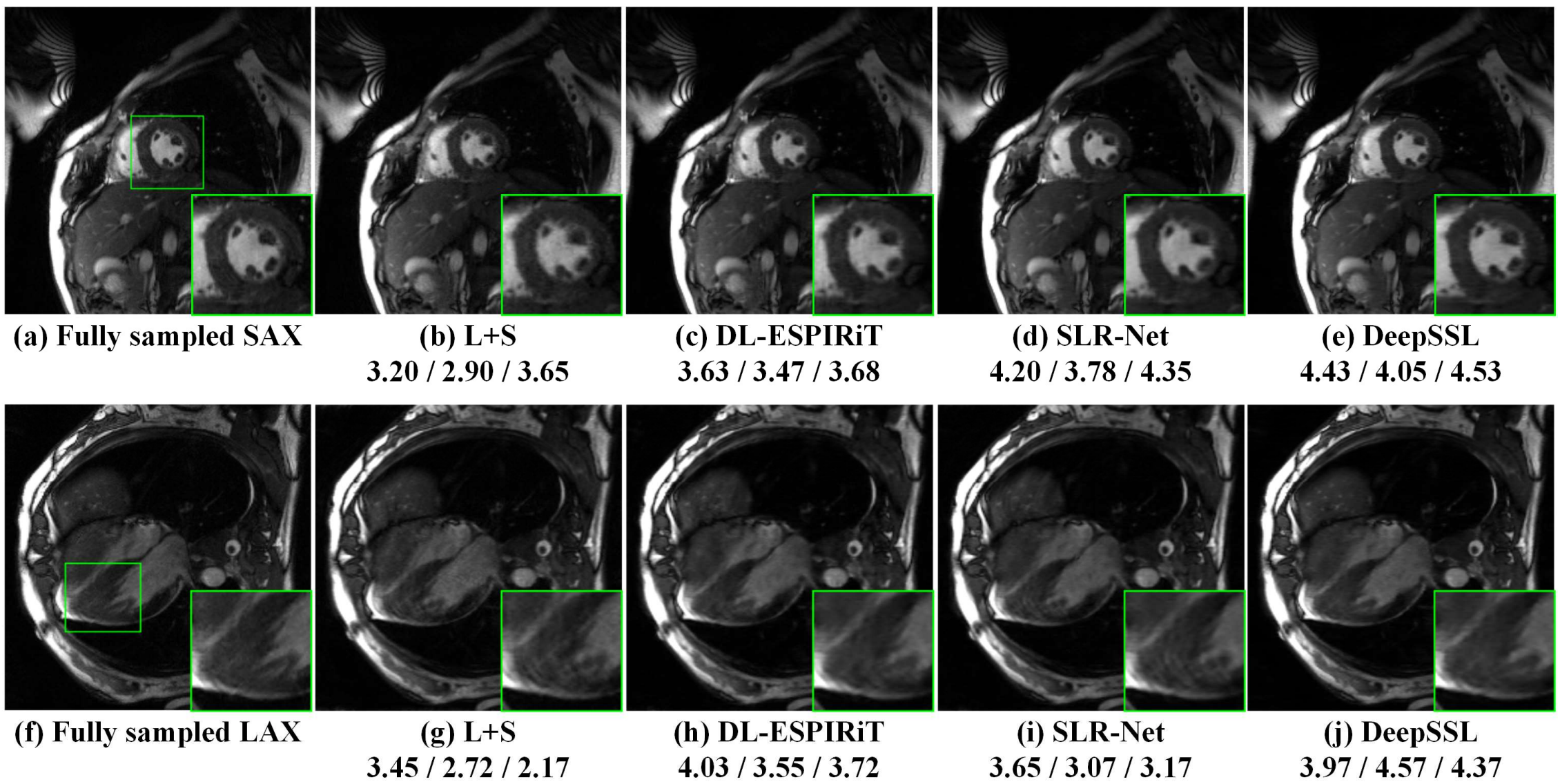

Fig. 9. Reconstruction results of a cardiac cine patient dataset of SAX and LAX views using different methods. (a) (or (f)) is the fully sampled image with the green zoom-in window. (b)-(e) (or (g)-(j)) are reconstructed images with zoom-in details. Note: The k-t random undersampling pattern with AF=6 is used. Each image lists the mean scores of six readers in the signal-to-noise ratio/artifacts suppression/overall image quality.

task of cardiac segmentation, which played a crucial role in clinical diagnosis and analysis.

We first trained a state-of-the-art cardiac segmentation network [58] using a public CMRxRecon dataset with manual segmentation ground truth [28]. Then, this network was used to compare the performance of reconstructed images using different methods for SAX cardiac segmentation. To quantify the segmentation accuracy, the Dice coefficient was utilized. A higher Dice coefficient indicates the better segmentation.

TABLE IV

DICE COEFFICIENT (×10$^{-2}$) IN THE CARDIAC SEGMENTATION [MEAN±STD].

| Method | LV | MYO | RV | Average |
|---|---|---|---|---|
| Fully sampled | **94.93±2.71** | **84.93±3.95** | **93.92±3.22** | **91.26±4.49** |
| DCCNN | 94.30±2.75 | 82.46±4.79$^{\dagger}$ | 92.87±3.22 | 89.88±5.28$^{\dagger}$ |
| DL-ESPIRiT | 94.42±2.73 | 82.72±4.59 | 92.49±3.22$^{\dagger}$ | 89.88±5.12$^{\dagger}$ |
| SLR-Net | 94.23±2.61$^{\dagger}$ | 82.54±4.63$^{\dagger}$ | 92.70±3.23 | 89.83±5.19$^{\dagger}$ |
| DeepSSL | **94.55±2.63** | **83.34±4.49** | **93.41±3.22** | **90.43±5.04** |

Note: Images for segmentation are reconstructed using different methods under the k-t random undersampling pattern with AF=6. The means and standard deviations are computed over 100 images of healthy cases, respectively. "LV", "MYO", "RV" denote left ventricle, myocardium, and right ventricle, respectively. "Average" denotes averaging the values of the three Dice coefficients. The Dice coefficients of fully sampled and the closest value to that of fully sampled are bold faced. "†" means the compared method has statistically differences ($p<0.05$) compared to our DeepSSL under Wilcoxon signed-rank test.

Table IV shows that the improvements of our DeepSSL in the average Dice coefficient of three segmentation regions are statistically difference, and it achieves better cardiac segmentation than other compared methods ($p<0.05$). Meanwhile, DeepSSL's results are closest to the fully sampled ones. Figs. 10(a)-(f) presents a typical case of segmentation results to show that all methods can achieve comparable segmentation when their reconstructed cardiac edges are relatively clear. Furthermore, our better edge reconstruction of the right ventricle region benefits more accurate segmentation in some cases (Figs. 10(g)-(l)).

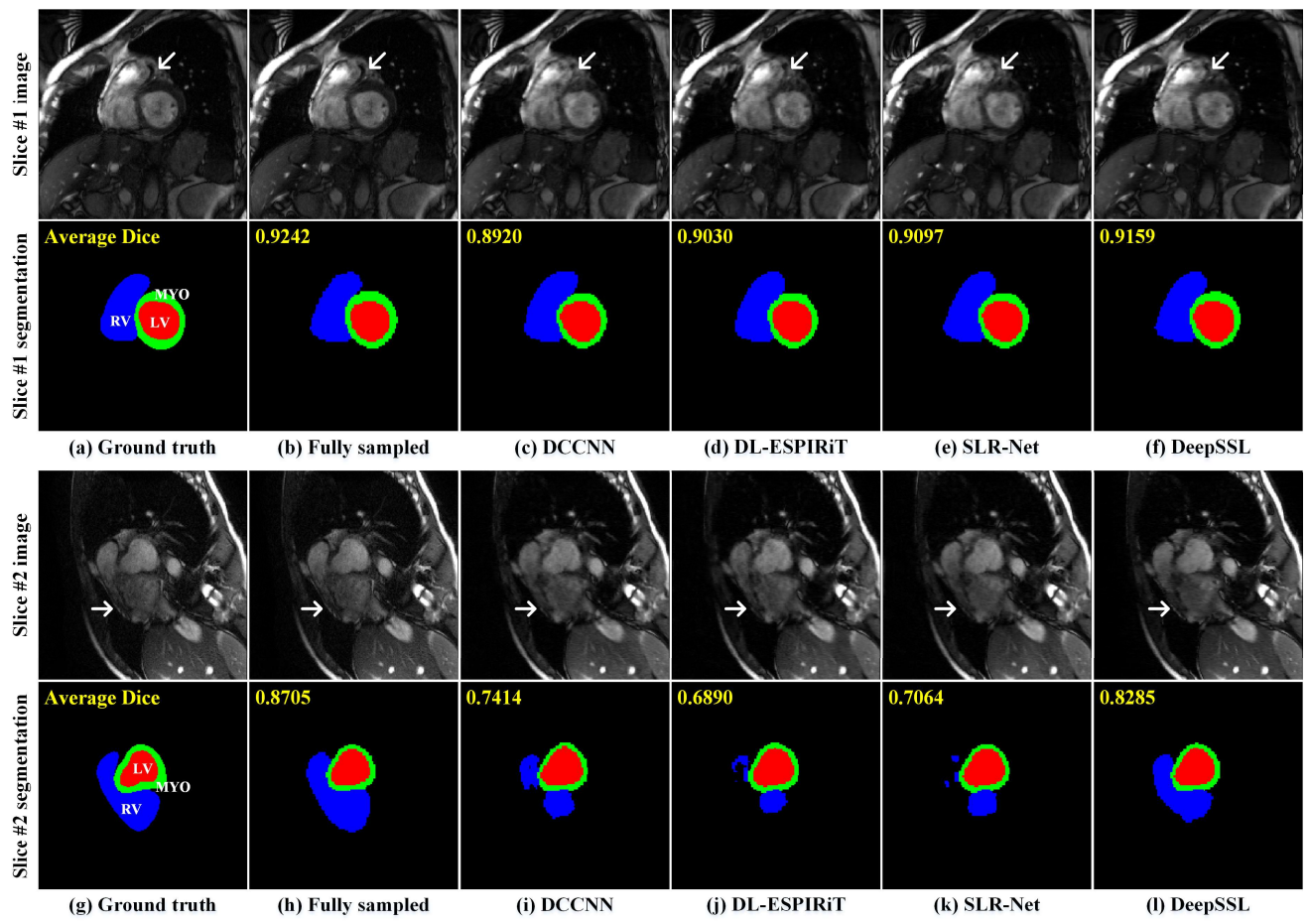

Fig. 10. Cardiac reconstruction and segmentation results of different methods. (a) (or (g)) is the fully sampled image and segmentation ground truth. (b) (or (h)) is the fully sampled image and segmentation result of fully sampled images. (c)-(f) (or (i)-(l)) are segmentation results from different reconstructed images. Note: The k-t random undersampling pattern with AF=6 is used. Edges of right ventricle are marked with white arrows.

### G. Prospective Undersampling: Real-Time MRI

In addition to the retrospective studies, we further explored the robustness of the proposed method to the unseen prospective undersampling scenario.

A free-breathing undersampled cardiac cine data from the public OCMR dataset [33] was used. It has 3 SAX slices of size 320×120 with 65 temporal frames and is prospectively undersampled using a k-t VISTA pattern with AF=8. In this experiment, all deep learning methods were trained under k-t VISTA undersampling with AF=8 (Fig. 11(a)) but used a same dataset as before.

Fig. 11 shows that the proposed DeepSSL is superior to other compared methods in terms of artifacts suppression and details preservation, and captures the organ motion precisely. L+S, DCCNN, and DL-ESPIRiT still exhibit obvious streaking artifacts (Marked with white arrows) and spatial blurring. SLR-Net provides nice dynamic information, but spatial artifacts remain slightly.

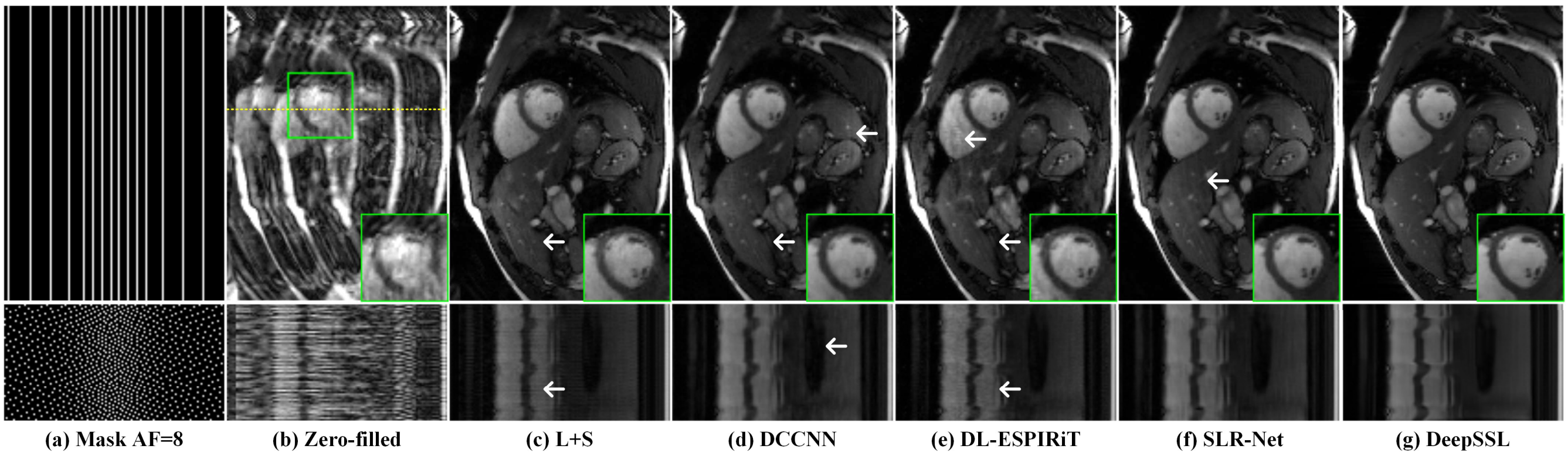

Fig. 11. Prospective undersampling reconstruction results of a real-time cardiac cine dataset using different methods. (a) is the k-t VISTA pattern with AF=8. (b) is the zero-filled image with the green zoom-in window. (c)-(g) are reconstructed images with zoom-in details. Note: The 2D spatiotemporal images are given for each reconstruction and their corresponding extracted position is marked by the yellow dotted line. Some obvious artifacts are marked with white arrows.

## H. Ablation Study

To evaluate the contribution of deep spatial sparse module and deep temporal low-rank module to reconstructions, we compared the proposed DeepSSL with its two variants that only impose spatial or temporal priors, named as Spatial-only and Temporal-only, respectively.

Fig. 12 shows that Spatial-only network removes most spatial artifacts but remains strong temporal artifacts, while Temporal-only network captures the dynamic information more precisely and also recovers most spatial structure. The proposed DeepSSL with spatiotemporal priors outperforms other two variants both visually and quantitatively. These results show the effectiveness of utilizing two complementary priors in dynamic MRI reconstruction.

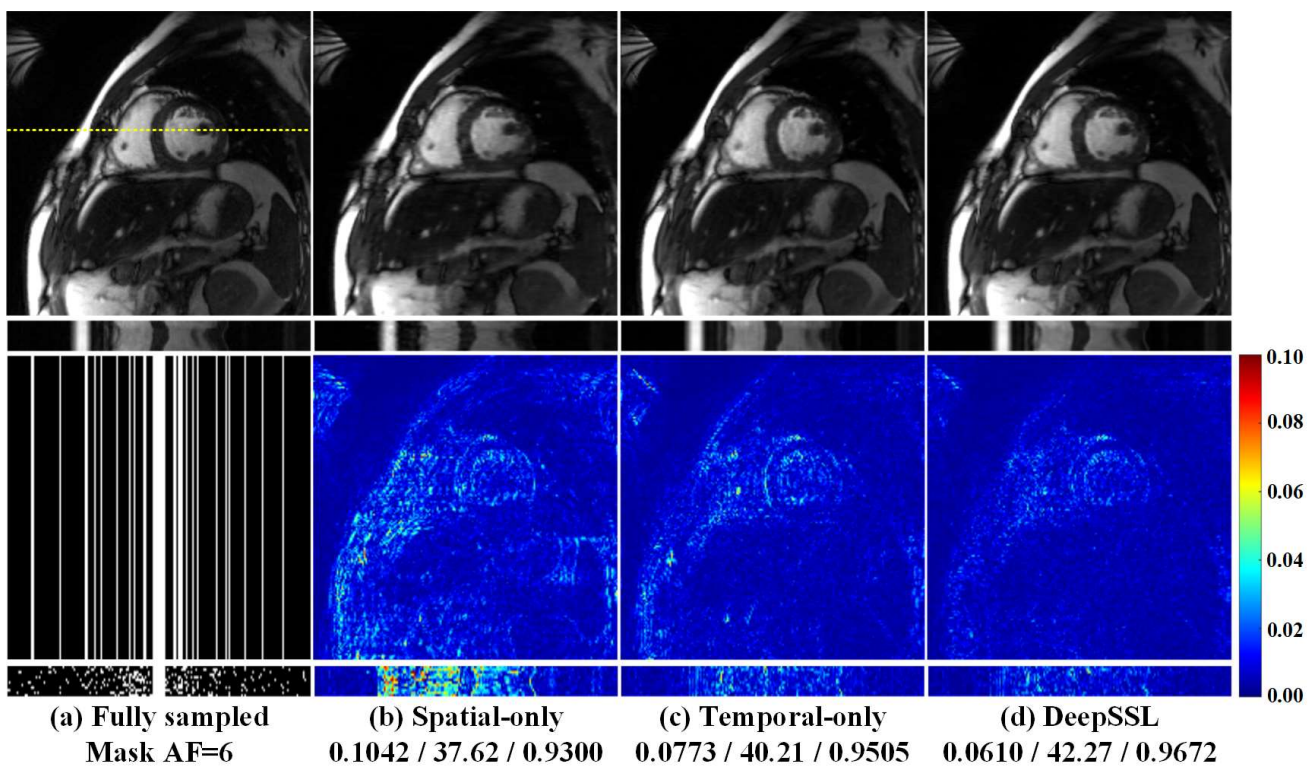

Fig. 12. Ablation study. (a) is the fully sampled image and the k-t random undersampling pattern with AF=6. (b)-(d) are reconstructed images and the corresponding error maps. Note: The 2D spatiotemporal images are given for each reconstruction and their corresponding extracted position is marked by the yellow dotted line. RLNE/PSNR(dB)/SSIM are listed for each reconstruction.

## I. Spatial Constraint Compensation

To compensate the performance loss due to the lack of spatial constraints along the FE direction in separable learning which is mentioned in Section IV-C, we offered a possible idea for adding spatial constraints.

Here, the FE-direction finite difference constraint was added in the network reconstruction stage, called DeepSSL-SC (Fig. 13), to enhance the correlation between rows of spatial image and compensate spatial smoothness. Due to the simplicity of the finite difference operator, it can improve image quality, especially in SSIM (Fig. 14), without increasing computational burden and any re-training.

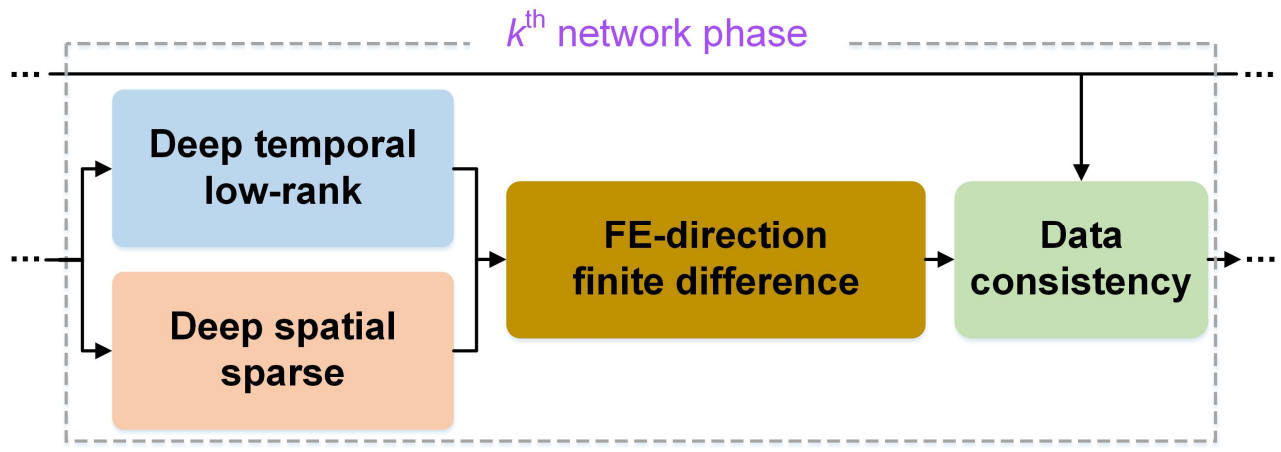

Fig. 13. The network architecture of the proposed method with the spatial constraint in the reconstruction stage. The FE-direction finite difference constraint is added for enhancing spatial image smoothness without the need for any re-training.

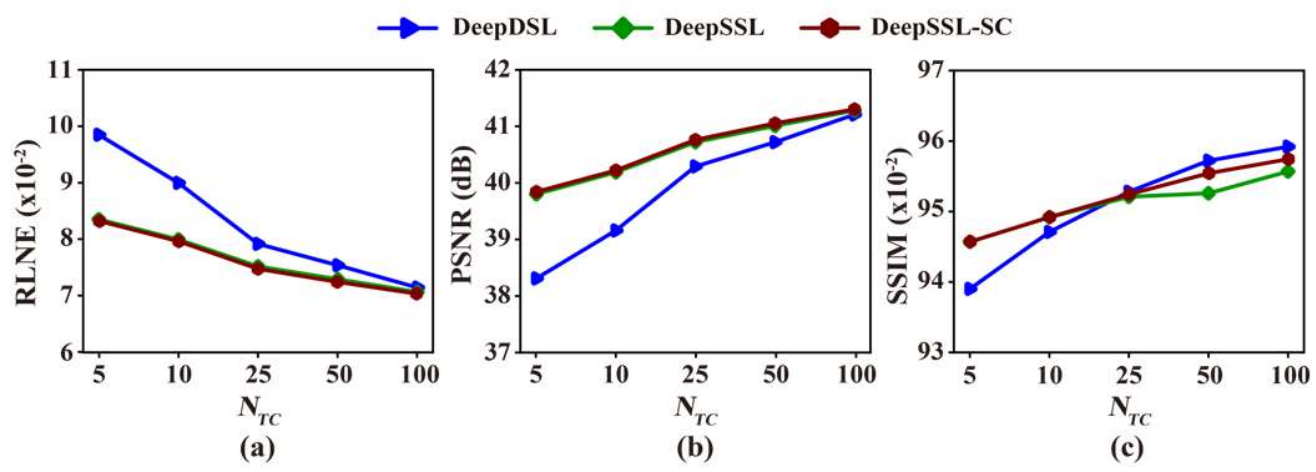

Fig. 14. Quantitative comparison of reconstructions of DeepDSL, DeepSSL, and DeepSSL-SC using different number of training cases $N_{TC}$ of a cardiac cine dataset. (a)-(c) are the mean of RLNE, PSNR, SSIM of the reconstructions of the SAX view, respectively. Note: The k-t random undersampling pattern with AF=6 is used. The means are computed over all 10 test data of healthy cases.

# V. Discussions

In this work, we propose a dimension-reduced separable learning scheme for the high-dimensional problem of dynamic cardiac MRI reconstruction. This scheme focuses on 2D spatiotemporal data in the phase encoding-temporal (PE-t) space, making network training and generalization easier. By integrating the spatiotemporal priors, our network behavior becomes traceable, enhancing its interpretability and reliability.

The impressive benefits of our DeepSSL are its efficiency under highly limited training data and adaptability to patients and prospective undersampling. To reduce training data reliance, alternative approaches include transfer learning [59], model adaptation [60], physics-driven synthetic data learning [22, 61], and scan-specific reconstruction [62]. Unlike their reliance on large-scale pre-trained models for fine-tuning or online target training, our idea follows the separability of k-t undersampling to fully use available data, which is more intuitive and easier to follow.

In the future, we anticipate that the following aspects are worth exploring to address the limitations of this work:

1) Spatial constraints compensation. Considering the performance loss in separable learning due to the lack of spatial constraints along the FE, expect using finite difference in Section IV-I, we can further integrate other spatial smoothness [2, 10] and/or k-space self-consistency [14, 61]. However, the coming issues of regularization parameter tuning and increased computational time need to be carefully addressed.

2) Generalizable learning techniques. We attempt to employ emerging generative model [60, 63] with domain adaptation and/or physics-informed synthetic data learning [22, 61] for multi-center and multi-vendor generalization.

3) Non-Cartesian imaging. To extend our method to handle non-Cartesian data, non-uniform fast Fourier transform could be used to replace the forward encoding operator [64], and then use our direct version (DeepDSL) or separately process the spatial image's rows and columns using our separable version (DeepSSL).

4) Challenging dynamic conditions. The current strategy primarily involves quality control by radiologists to exclude tricky data with failed breath-hold and arrhythmias [34, 35]. Besides, we can also design a correction module to mitigate motion artifacts and use free-breathing real-time imaging for patients [65], instead of the current breath-hold pipeline.

## VI. Conclusion

Our work presents a novel dimension-reduced separable learning scheme, integrated with spatiotemporal priors, culminating in the development of a Deep Separable Spatiotemporal Learning network (DeepSSL) tailored for dynamic cardiac magnetic resonance imaging (MRI).

Extensive evaluations on *in vivo* datasets showcase that for both healthy cases and patients, the proposed DeepSSL achieves state-of-the-art performance, both visually and quantitatively, while reducing the demand for training cases by up to 75%. The reliability of DeepSSL in the diagnostic perspective is affirmed through validation by four experienced radiologists and two cardiologists. Additionally, our method proves beneficial for cardiac segmentation and exhibits robustness in prospectively undersampled data.

Our study demonstrates that separable deep learning effectively mitigates problem dimensions, reduces the reliance on training data volume, and lessens the computing device requirements. This innovative approach holds promise in addressing the escalating demand for high-dimensional data reconstruction in MRI applications.

## Acknowledgments

The authors thank Drs. Michael Lustig, Ricardo Otazo, Jo Schlemper, Christopher M. Sandino, Dong Liang, Yongchao Xu, Thomas Küstner, and Yue Hu for sharing their codes online.